\documentclass[11pt]{biophys_clean}

\usepackage{helvet,times}
\usepackage{bm,textcomp}
\usepackage{amssymb}

\jno{kxl014} 
\gridframe{N}
\cropmark{N}

\usepackage{amsmath}

\usepackage[round,numbers,sort&compress]{natbib}

\newcommand{\e}{{\rm e}}

\newcommand{\kb}{k_\mathrm{B}}

\usepackage[textsize=scriptsize]{todonotes}
\usepackage{mathtools}

\usepackage[colorlinks=true,linkcolor=cyan!50!blue!70!black!74,citecolor=cyan!50!blue!70!black!74,breaklinks=true]{hyperref}

\begin{document}

\setcounter{page}{1} 
\title{Thermodynamic bounds on the ultra- and infra-affinity of Hsp70 for its substrates}
\author{Basile Nguyen $^{1,2}$, David Hartich $^1$, Udo Seifert $^1$ and Paolo De Los Rios $^2$}
\address{$^1$II. Institut f{\"u}r Theoretische Physik, Universit{\"a}t Stuttgart, Stuttgart, Germany; $^2$ Laboratory of Statistical Biophysics, Institute of Physics, School of Basic Science and Institute of Bioengineering, School of Life Sciences, \'{E}cole Polytechnique F\'{e}d\'{e}rale de Lausanne (EPFL), Lausanne, Switzerland}

\begin{abstract}
{
The 70 kDa Heat Shock Proteins Hsp70 have several essential functions in living systems, such as protecting cells against protein aggregation, assisting protein folding, remodeling protein complexes and driving the translocation into organelles. These functions require high affinity for non-specific amino-acid sequences that are ubiquitous in proteins. It has been recently shown that this high affinity, called \textit{ultra-affinity}, depends on a process driven out of equilibrium by ATP hydrolysis. Here we establish the thermodynamic bounds for ultra-affinity, and further show that the same reaction scheme can in principle be used both to strengthen and to weaken affinities (leading in this case to infra-affinity). We show that cofactors are essential to achieve affinity beyond the equilibrium range. Finally, biological implications are discussed.}
{}{}%
\end{abstract}

\maketitle 

\section*{INTRODUCTION}
Most proteins must fold into specific three-dimensional structures (native states) to be functional and take part in cellular processes. During, and right after, translation, newly synthesized polypeptides are not yet fully folded. As a consequence, they still expose hydrophobic surfaces, that could lead to inter-protein interaction and cytotoxic aggregation \cite{chit06}. Furthermore, mutations or environmental cues, such as heat-shock or oxidative stress, can destabilize native proteins, leading to their unfolding, misfolding and potential aggregation. In cells, the protein quality control system acts to maximize the reliability of protein folding, and to clear proteins that cannot be driven back to their native state \cite{yuji13}. Defects in protein quality control are associated with age-related diseases such as type II diabetes, heart diseases, specific cancers and, most notably, neurodegenerative disorders (e.g Alzheimer's or Parkinson's diseases) \cite{labb15}.

Chaperones proteins are key players in protein quality control, and are present in all organisms. Their broadly recognized role is to assist the folding process, and minimize protein aggregation. Intriguingly, the action of most chaperones stringently depends on ATP hydrolysis, although in most cases its precise role has not been fully understood. Central among proteins is the 70 kDa Heat Shock Protein (Hsp70). Hsp70 is possibly the most versatile of the chaperones and takes part in disparate functions beyond quality control. It drives the translocation of hundreds of different proteins into mitochondria and the \textit{endoplasmic reticulum}, disassembles functional oligomers and facilitates protein translation, among others \cite{yuji13,fink16}. In order to be functional, Hsp70s must be able to strongly bind to a diverse array of amino-acid sequence. 

The structure of Hsp70 comprises two domains: the nucleotide binding domain (NBD), where ATP or ADP are lodged, and the substrate binding domain (SBD), which is made of two halves and is responsible for the interactions between Hsp70s and their substrates \cite{kity15}, see Fig.~\ref{fig:4state} for an illustration. In the ATP-bound state, the two halves of the SBD are preferably docked onto the NBD ("open" conformation, Fig.~\ref{fig:4state}), whereas in the ADP bound state the two halves of the SBD preferably detach from the NBD and bind to each other, forming a "closed" clamp (which remains linked to the NBD by a flexible linker). Note that since binding and unbinding rates are significant in the ADP bound state, spontaneous opening and closing occurs \cite{maye00a}. The spontaneous ATPase rate of Hsp70 is very low ($10^{-4}-10^{-3}$ $\text{s}^{-1}$), but is greatly accelerated (up to 1 $\text{s}^{-1}$) upon substrate binding and an associated, mandatory, J-domain containing protein \cite{kamp10}. Upon contact, thus, Hsp70 latches onto the substrate after rapid ATP hydrolysis, entrapping it into the closed clamp (Fig.~\ref{fig:4state}). Remarkably, the measured substrate affinity of the ATP-bound, open conformation is only slightly lower than the substrate affinity of the closed ADP-bound conformation ($K_\mathrm{D}^\textrm{ATP}> K_\mathrm{D}^\textrm{ADP}$, see Refs.~\cite{pier98,zuid13} or Table~\ref{4state_realrate} for experimental values).

Experiments, though, have shown that substrate binding occurs mainly in the ATP state rather than ADP state, despite the latter being the state characterized by the smallest dissociation constant \cite{lauf99,witt03}. It had been proposed that this effect was inherently due to the non-equilibrium, ATP-consuming, nature of Hsp70s. Recently, this enhanced affinity (dubbed \textit{ultra-affinity}) was linked to the kinetic properties of the ATP-bound and ADP-bound states \cite{rios14}. The substrate binding and unbinding rates are faster for ATP-bound Hsp70s, because the SBD is preferably open and easily accessible, than for ADP-bound Hsp70s, whose SBD is preferably closed and thus difficult to bind to, but also difficult to unbind from (see Fig.~\ref{fig:4state}). Due to an excess of ATP in living cells and in the vast majority of experiments, most free Hsp70 molecules are bound to ATP. As a consequence, substrate binding occurs mostly in the ATP-bound Hsp70, with the fastest binding rate $k^\textrm{ATP}_+$. ATP hydrolysis rapidly follows, with the closure of the SBD on the substrate. Substrate unbinding takes place then with the smallest dissociation rate, $k^\textrm{ADP}_-$. The effective non-equilibrium dissociation constant can thus be lowered down to $K_\textrm{eff} = k^\textrm{ADP}_-/k^\textrm{ATP}_+$, which is not related to the individual dissociation constants of the ATP-bound and ADP-bound states. It can be smaller than the dissociation constant attainable without hydrolysis, which is the average of the dissociation constants of the two nucleotide-bound states, weighted by their respective populations. Ultra-affinity is a remarkable principle, which allows Hsp70s to bind very effectively to a broad, non-specific, array of amino-acid sequences. Experiments show that Hsp70s bind to their substrates with different intrinsic affinities \cite{pier98}. When energy from ATP hydrolysis is available, Hsp70s can bind to their substrates with a higher affinity which might be further enhanced during stresses. We provide new insight into this selectivity mechanism in response to stress. Furthermore, we show that Hsp70 can achieve such a selectivity depending on stress level, which relates to heat shock experiments that show changes in nucleotide levels \cite{soin05,lill84,find83} and changes in the activity of cofactors \cite{sieg06,grim01} during stress. 

 A careful analysis of ultra-affinity, though, reveals that the energy budget of the process should also be taken into account. In this work, we consider a thermodynamic description of the Hsp70 system. Specifically, we characterize the relation between affinity and energy consumption by computing the thermodynamic bounds of ultra-affinity. Moreover, we show that it is possible to obtain the opposite of ultra-affinity, namely \textit{infra-affinity}, that is an affinity which is lower than what would be possible at equilibrium. Note that the Hsp70 system shares many similarities with kinetic proofreading \cite{hopf74,nini75,benn79,gasp16} where error reduction is achieved with a chemical force driving the system out of equilibrium \cite{ehre80,qian06,qian07,muru12,muru14}. A similar ultra-sensitive response was found for the E. coli chemotaxis system \cite{tu08a} where the increase of sensitivity by a non-equilibrium driving force was described and compared with kinetic proofreading \cite{hart15}.

\section*{METHODS: Local detailed balance}
\begin{figure}
\centering
\includegraphics[width=.5\columnwidth]{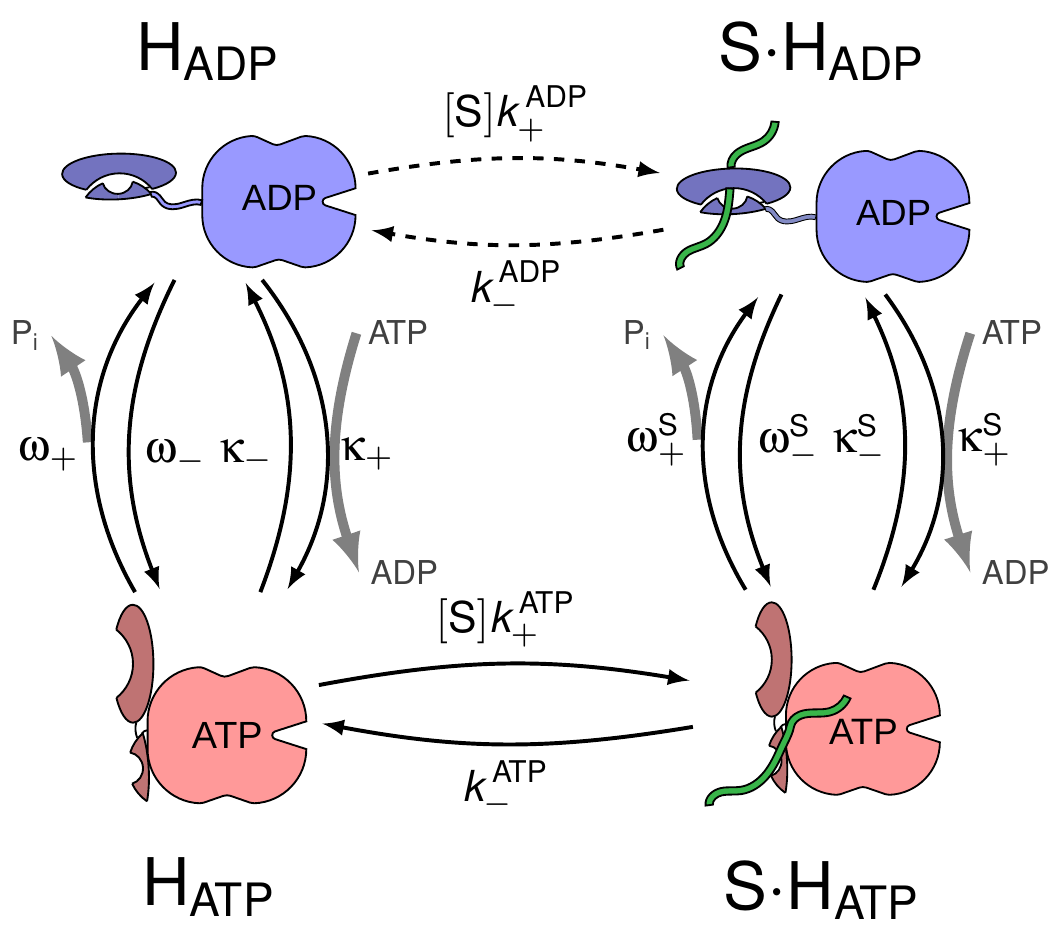}
\caption{Canonical model of the Hsp70 cycle. Horizontal rates correspond to binding/unbinding of a substrate with Hsp70's two states. The ADP-state has a lower dissociation constant and slow binding kinetics $k_\pm^{\text{ADP}}$. The ATP-state has a higher dissociation constant and fast binding kinetics ($k_\pm^{\text{ATP}} > k_\pm^{\text{ADP}}$). Vertical rates correspond to hydrolysis $\omega_{\pm}$  and nucleotide exchange $\kappa_{\pm}$ reactions. Specifically, $\omega_+$ is a release of $\text{P}_\text{i}$ and $\kappa_+$ is an exchange of ADP with ATP, both of which are emphasized by thick arrows.}
\label{fig:4state}
\end{figure}

We model the Hsp70 system by a canonical four state system from \cite{rios14}, see Fig.~\ref{fig:4state} for an illustration. The Hsp70 system can be in an open ATP state ($\text{H}_\text{ATP}$, $\text{S}\cdot\text{H}_\text{ATP}$) or in a closed ADP state ($\text{H}_\text{ADP}$, $\text{S}\cdot\text{H}_\text{ADP}$), where ``S'' labels the presence of a substrate. Chemical forces arising from an ATP hydrolysis cycle drive the system out of equilibrium, which allow Hsp70 to tune its affinity to substrates. To better understand the benefit of such chemical forces, we have to explain the local detailed balance relation \cite{seif11}, which connects the dynamics of single reactions with the laws of thermodynamic.

We introduce the local detailed balance relation by considering the individual reactions corresponding to substrate binding and unbinding, which is illustrated in Fig.~\ref{fig:4state} by horizontal transitions. The binding and unbinding of the substrate to and from the chaperones corresponds to the reaction
\begin{equation}
  \mathrm{H}_X+\mathrm{S}\xrightleftharpoons[{k_-^X}]{[\mathrm{S}]k_+^X}\mathrm{S}\cdot\mathrm{H}_X,
  \label{eq:substrate}
 \end{equation}
where $X=\mathrm{ATP},\mathrm{ADP}$ indicates the state of the heat shock protein and $[\mathrm{S}]k^X_+,k^X_-$ are transition rates. At temperature $T$,  the transition rates must satisfy the local detailed balance relation \cite{seif11}, which for $X=\mathrm{ATP},\mathrm{ADP}$ reads
\begin{equation}
 \kb T\ln \frac{[\mathrm{S}]k_+^X}{{k_-^X}}=F_{\mathrm{H}_X}-F_{\mathrm{S}\cdot\mathrm{H}_{X}}+\mu_\mathrm{S},
 \label{eq:LDB_substrate}
\end{equation}
where $F_{\mathrm{H}_X}$ is the free energy of state $\mathrm{H}_X$,
$F_{\mathrm{S}\cdot\mathrm{H}_{X}}$ the free energy of state $\mathrm{S}\cdot\mathrm{H}_{X}$, $\mu_\mathrm{S}$ the chemical potential of the substrate, and $\kb$ Boltzmann's constant. Experimental measurements of binding rates for the Hsp70 system can be found in Table~\ref{4state_realrate}.

\begin{table}
 \centering
  \begin{tabular}{|c|l|}
  \hline
  $k^\textrm{ADP}_+$ & $10^{-3}\, \text{s}^{-1}\mu\text{M}^{-1}$ \cite{russ98,they96} \\
   $k^\textrm{ADP}_-$ & $4.7\cdot 10^{-4}\, \text{s}^{-1}$  \cite{russ98,they96}\\
    $k^\textrm{ATP}_+$ & $4.5\cdot 10^{-1}\, \text{s}^{-1}\mu\text{M}^{-1}$  \cite{schm94,gisl98}  \\
     $k^\textrm{ATP}_-$ &$2\, \text{s}^{-1}$ \cite{schm94,gisl98}  \\
     $K^\textrm{ADP}_\text{D}$ & $0.47\,\mu\text{M}$\\
        $K^\textrm{ATP}_\text{D}$& $4.4\,\mu\text{M}$\\
     \hline
  \end{tabular}
  \caption[Biological relevant parameters]{\label{4state_realrate}Experimental transition rates for the Hsp70 system and dissociation constants $K^\textrm{ADP}_\text{D}=k^\textrm{ADP}_-/k^\textrm{ADP}_+$, $K^\textrm{ATP}_\text{D}=k^\textrm{ATP}_-/k^\textrm{ATP}_+$.}
\end{table}

The vertical transitions in Fig.~\ref{fig:4state} involve the consumption of chemical energy due to ATP hydrolysis allowing the system to outperform equilibrium chaperone systems. More precisely, the vertical transition in Fig.~\ref{fig:4state} correspond to a hydrolysis reaction
 \begin{equation}
  \mathrm{H}_\mathrm{ATP}\xrightleftharpoons[{\omega_-}]{\omega_+}\mathrm{H}_\mathrm{ADP}+\mathrm{P_i},
  \label{eq:hydrolysis}
 \end{equation}
 and a nucleotide exchange reaction
 \begin{equation}
  \mathrm{H}_\mathrm{ADP}+\mathrm{ATP}\xrightleftharpoons[{\kappa_-}]{\kappa_+}\mathrm{H}_\mathrm{ATP}+\mathrm{ADP}
  \label{eq:exchange},
 \end{equation}
 where $\kappa_\pm$ and $\omega_{\pm}$ are transition rates in the absence of a substrate. Analogously, in the presence of a substrate the transition rates are denoted by an additional superscript ``$S$'', see $\kappa_\pm^S$ and $\omega_{\pm}^S$ in Fig.~\ref{fig:4state}. Note that in reality, the nucleotide exchange is a two-step reaction involving unbinding of ADP (ATP) and binding of ATP (ADP). Nevertheless, Eq.~\ref{eq:exchange} is an effective relation which is equivalent as nucleotide binding is very fast and nucleotide affinity with Hsp70 is high. Performing one step in ``+''-direction of reaction \ref{eq:hydrolysis}  and then one step in ``+''-direction of reaction \ref{eq:exchange} does not change the state of the Hsp70 system, whereas it turns one ATP into an ADP and $\mathrm{P_i}$. Such a complete cycle consumes a chemical energy (work)
 \begin{equation}
  \varDelta\mu\equiv\mu_\mathrm{ATP}-\mu_\mathrm{ADP}-\mu_\mathrm{P_i}=\varDelta\mu_0+\kb T\ln\frac{[\mathrm{ATP}]}{[\mathrm{ADP}][\mathrm{P_i}]},
  \label{eq:def_dmu}
 \end{equation}
where $\mu_X$ is the chemical potential of species $X=\mathrm{ATP},\mathrm{ADP},\mathrm{P_i}$. The last equality in Eq.~\ref{eq:def_dmu} is the approximation for an ideal solution, where $[X]$
is the concentration of species $X=\mathrm{ATP},\mathrm{ADP},\mathrm{P_i}$, and $\varDelta\mu_0$ a reference value. Equilibrium corresponds to $\varDelta\mu=0$, whereas under physiological conditions an excess of ATP is maintained that implies $\varDelta\mu>0$. For such a cycle the local detailed balance relation implies
 \begin{equation}
  \beta\varDelta\mu=\ln\frac{\kappa_+\omega_+}{\kappa_-\omega_-}=\ln\frac{\kappa_+^S\omega_+^S}{\kappa_-^S\omega_-^S},
   \label{eq:LDB}
 \end{equation}
 where $\beta=1/(\kb T)$ is the inverse thermal energy. This relation connects the kinetics of hydrolysis Eq.~\ref{eq:hydrolysis} and nucleotide exchange Eq.~\ref{eq:exchange} to the chemical driving force $\varDelta\mu$ from Eq.~\ref{eq:def_dmu}. Along a complete ATP hydrolysis cycle, this chemical energy $\varDelta\mu$ is dissipated in the environment. Moreover, a constant chemical driving force $\varDelta\mu>0$ (supply of ATP) drives the chaperone system
 into a non-equilibrium steady state, leaving more room to tune the system compared to a system with an equilibrium Boltzmann distribution, where $\varDelta\mu=0$.

\section*{RESULTS}
We are interested in a thermodynamic relation between energy consumption and Hsp70's affinity for its substrates. We consider the effective dissociation constant $K_\textrm{eff}$ which measures how well Hsp70s can bind to their substrates. It is defined as
\begin{equation}
K_\textrm{eff} = \textrm{[S][Hsp70]/[Hsp70}\cdot\textrm{S]}
\end{equation} 
 where [S] is the concentration of free substrate, [Hsp70] is the concentration of free Hsp70s and [Hsp70$\cdot$S] is the concentration of substrates bound to Hsp70 (see Appendix A for the full expression). In equilibrium, the dissociation constant is a linear combination of the ADP and ATP dissociation constants, therefore, $K_\mathrm{D}^\textrm{ADP}\le  K_\text{eff}^\text{eq}\le K_\mathrm{D}^\text{ATP}$. Under physiological conditions, an excess of ATP is maintained which induces a positive chemical force ($\varDelta\mu>0 $). Under these non-equilibrium conditions, it has been found that $K_\textrm{eff} < K_\mathrm{D}^\textrm{ADP}$ can be achieved, which is called ultra-affinity \cite{rios14}. We provide a thermodynamic description of ultra-affinity and show that this system can also achieve infra-affinity for its substrates.

\begin{figure}
\centering
\includegraphics[width=.7\columnwidth]{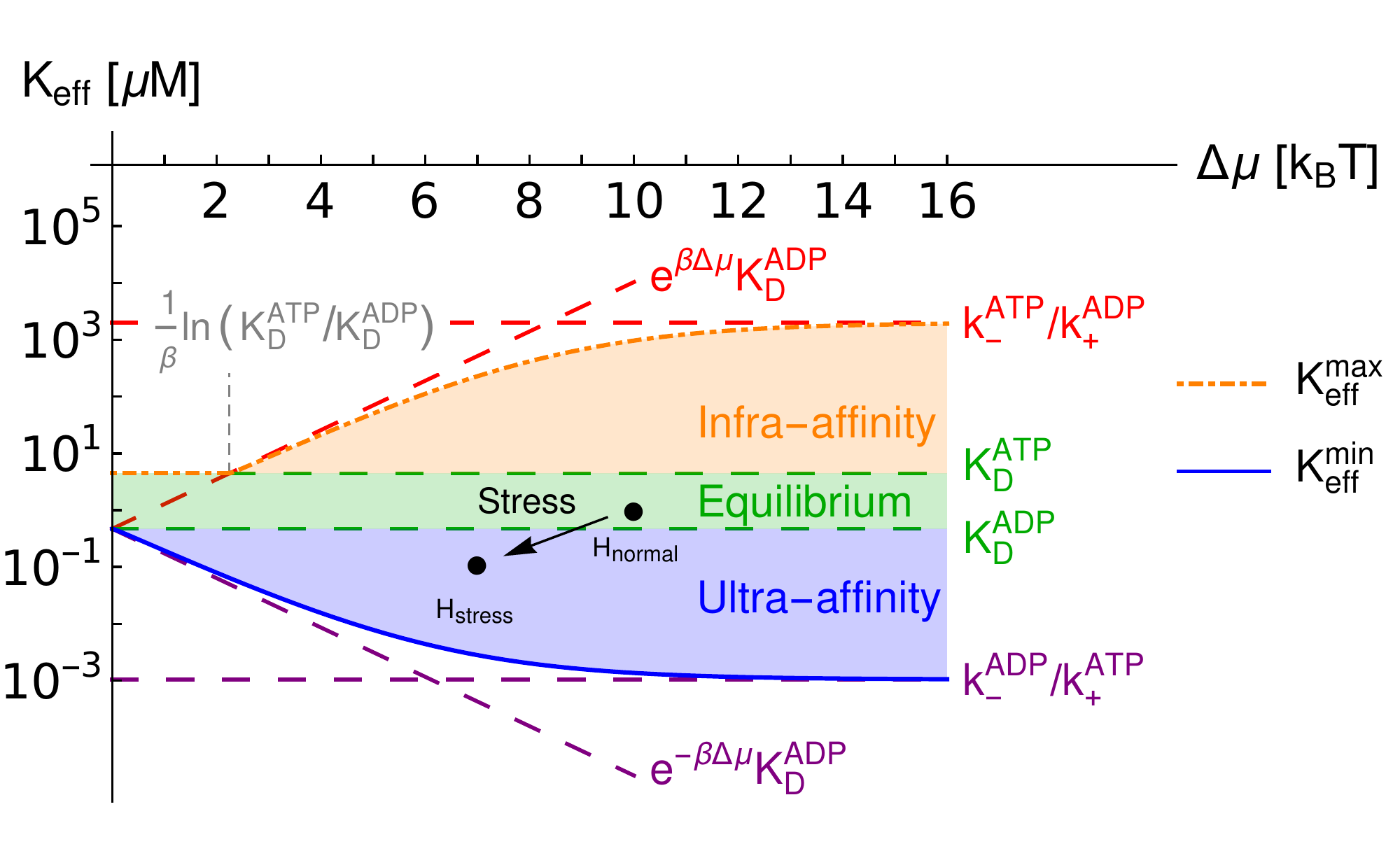}
\caption{Thermodynamic bounds on the effective dissociation constant based on experimental binding rates as listed in Table~\ref{4state_realrate}. In equilibrium, the dissociation constant is bounded by the ADP and ATP dissociation constant $K_\textrm{eff}^\text{ADP,ATP}$. Ultra-affinity corresponds to $K_\textrm{eff} < K_\mathrm{D}^\textrm{ADP}$ and infra-affinity to $K_\textrm{eff} > K_\mathrm{D}^\textrm{ADP}$, these regimes can be achieved only with a non-equilibrium driving force. In addition, infra-affinity can be achieved only with a minimum driving force $\varDelta\mu > (1/\beta)\ln (K_\mathrm{D}^\text{ATP}/K_\mathrm{D}^\text{ADP})$. The dissociation constant can be optimally tuned close to the ultra-affinity limit ($k^\textrm{ADP}_-/k^\textrm{ATP}_+$)  and the infra-affinity limit ($k^\textrm{ATP}_-/k^\textrm{ADP}_+$), provided sufficient driving force $\varDelta\mu$. The black arrow shows the effect of stress which can trigger ultra-affinity in response to experimental observations \cite{soin05,lill84,find83,sieg06,grim01} as explained in the main text. The normal system $\text{H}_\text{normal}$ has the following parameters ($\omega_+ = 10^{-4}\text{s}^{-1}, \omega_+^\text{S} = 0.01\text{s}^{-1}, \kappa_+=27\text{s}^{-1}, \kappa_- = 0.5\text{s}^{-1}$ and $\kappa_+^\text{S} = 0.003\text{s}^{-1}$). The other rates are computed using local detailed balance, see Eqs.~\ref{cond1}-\ref{cond3} in Appendix A.}
\label{fig:bound}
\end{figure}

We first consider ultra-affinity qualitatively as explained in \cite{rios14}. In the case of Hsp70, the substrate binding and unbinding kinetics is faster in the ATP state as it is in the ADP state (see Table~\ref{4state_realrate}). The slow substrate binding and unbinding kinetics in the ADP state is indicated by horizontal dashed arrows in Fig.~\ref{fig:4state}. For ultra-affinity, when a substrate is bound to Hsp70 it should ideally switch to the closed $\text{S}\cdot\text{H}_\text{ADP}$ configuration to benefit from slow substrate unbinding $k^\textrm{ADP}_-$, whereas in the absence of the substrate Hsp70 should ideally switch to the open $\text{H}_\text{ATP}$ configuration to benefit from fast substrate binding $k^\textrm{ATP}_+$. Infra-affinity on the other hand requires an opposite switching behavior. When a substrate is bound to Hsp70 it should ideally switch to the open $\text{S}\cdot\text{H}_\text{ATP}$ configuration to benefit from fast substrate unbinding $k^\textrm{ATP}_-$, whereas in the
absence of the substrate it should ideally switch to the closed $\text{H}_\text{ADP}$ configuration to benefit from slow substrate binding $k^\textrm{ADP}_+$. Note that with a finite budget of chemical energy $\varDelta\mu$, such an ideal switching behavior cannot be perfectly realized as shown in the following.

An optimization of the effective dissociation constant $K_\textrm{eff}$, while keeping the thermodynamic constraint Eqs.~\ref{eq:LDB_substrate} and \ref{eq:LDB} allow us to derive bounds for $K_\textrm{eff}$  as shown in Fig.~\ref{fig:bound} (see Appendix A for the derivation). For a given energy budget $\varDelta\mu$ and substrate kinetics $k_\pm^\mathrm{ATP},k_\pm^\mathrm{ADP}$ (see Table~\ref{4state_realrate} for experimental values), we optimize $K_\textrm{eff}$ with respect to the hydrolysis rates $\omega_{\pm},\omega_{\pm}^S$ and nucleotide exchange rates $\kappa_{\pm},\kappa_{\pm}^S$. The driving force $\varDelta \mu$ then determines the maximum decrease of dissociation constant allowing $K_\textrm{eff} < K_\mathrm{D}^\textrm{ADP}$. First, $e^{-\beta\varDelta\mu}K_\mathrm{D}^\text{ADP} \leq K_\textrm{eff}$ provides a simple lower bound. Second, $k^\textrm{ADP}_-/k^\textrm{ATP}_+\leq K_\textrm{eff}$ provides another simple lower bound, which is relevant in the limit of infinite driving ($\varDelta\mu \rightarrow \infty$). It corresponds to the ideal case where binding occurs only in the open ATP state and unbinding in the closed ADP state. Both lower bounds on the effective dissociation constant are shown as dashed purple lines in Fig.~\ref{fig:bound}. Finally, a minimization of the effective dissociation constant $K_\text{eff}$ by varying the kinetic parameters while satisfying the energetic constraints leads to an analytic lower bound $K_\text{eff}^{\min}$. The analytical expression and derivation of $K_\text{eff}^{\min}$ is presented in the Appendix A. Note that in the case of high substrate concentration, the effective dissociation constant $K_\text{eff}$ is bounded by the equilibrium dissociation constants $K_\mathrm{D}^\textrm{ADP,ATP}$. 

Infra-affinity ($K_\textrm{eff}>K_\mathrm{D}^\textrm{ATP}$), in contrast with ultra-affinity, requires investing a minimum free energy difference $\varDelta\mu > (1/\beta)\ln (K_\mathrm{D}^\text{ATP}/K_\mathrm{D}^\text{ADP})$ which must work against an equilibrium bias that arises from the allosteric interaction induced by different dissociation constants $K_\mathrm{D}^{\text{ADP, ATP}}$ (see Fig.~\ref{fig:bound}). Hence, the binding and unbinding kinetics of Hsp70 from Table~\ref{4state_realrate} favors ultra-affinity. For $\varDelta\mu > (1/\beta)\ln (K_\mathrm{D}^\text{ATP}/K_\mathrm{D}^\text{ADP})$, however, infra-affinity can be achieved, where a simple upper bound on the effective dissociation constant $K_\text{eff}$ is given by $\min \left[e^{\beta\varDelta\mu}K_\mathrm{D}^\text{ADP}, k^\textrm{ATP}_-/k^\textrm{ADP}_+ \right]$ which is indicated by two dashed red lines in Fig.~\ref{fig:bound}. A more detailed calculation, as shown in Appendix A, leads to an upper bound on infra-affinity $K_\text{eff}^{\max}$, which is obtained from a maximization of the effective dissociation constant $K_\text{eff}$ while keeping the energetic constraint fixed. Unlike ultra-affinity, the upper bound $k^\textrm{ATP}_-/k^\textrm{ADP}_+$ corresponds to the case, where binding occurs only in the closed ADP state and unbinding only in the open ATP state. Most remarkably, a similar concept was found in kinetic proofreading, where it was called "anti-proofreading" \cite{muru14}. This kinetic limit uses the non-equilibrium features to lower the discrimination between substrates. In this regime, the non-equilibrium force must also overcome a critical equilibrium bias. 

\begin{figure}
\centering
\includegraphics[width=1\columnwidth]{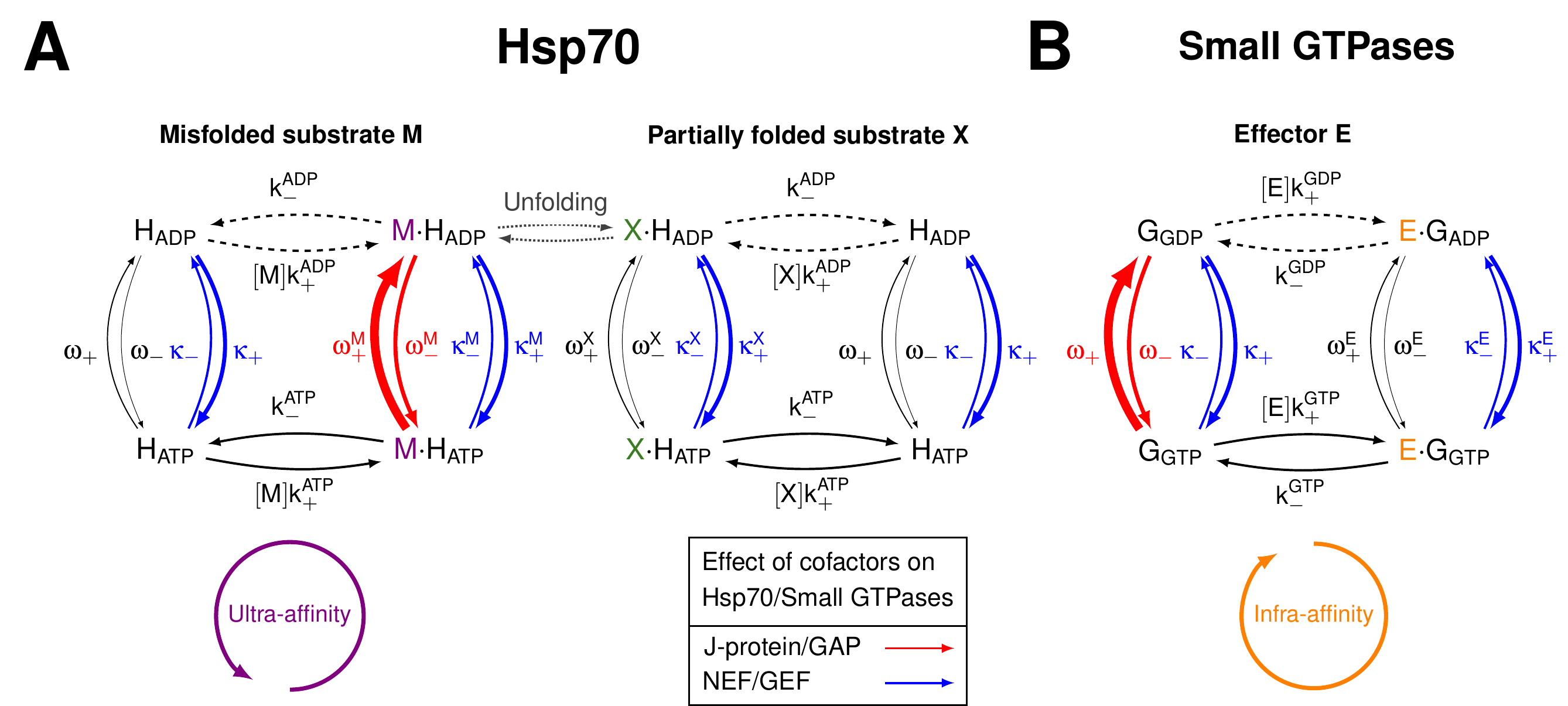}
\caption{The role of cofactors in Hsp70s and small GTPases. (A) The effect of Hsp70's cofactors depends on the state of the substrate. For misfolded substrate (left panel), J-proteins binds to misfolded substrate and strongly catalyze hydrolysis $\omega_\pm^\text{M}$ reactions \cite{kamp10,lauf99}. Nucleotide exchange factors (NEFs) catalyze the nucleotide exchange $\kappa_\pm, \kappa_\pm^\text{M,X}$ reactions. Ultra-affinity requires reaction cycling in the counter clockwise direction as shown in the lower left panel. For partially folded substrate (middle panel), J-proteins do not interact with them. However, the mechanism of NEFs does not depend on the state of substrate \cite{brac15}. (B) Small GTPases have cofactors which promote infra-affinity. Guanine nucleotide exchange factors (GEFs) are similar to NEFs, they facilitate GDP dissociation which boost nucleotide exchange $\kappa_\pm, \kappa_\pm^\text{E}$ reactions. Contrary to J-proteins, GTPase activating proteins (GAPs) compete with substrate effector to  bind small GTPases and catalyze the hydrolysis $\omega_\pm$ reactions \cite{cher13}. In addition, we have $K_\mathrm{D}^\text{GDP} > K_\mathrm{D}^\text{GTP}$ \cite{berg09,burg08,junu04}, where $K_\mathrm{D}^\text{GDP} = k_-^\text{GDP} /k_+^\text{GDP} $ and $K_\mathrm{D}^\text{GTP} = k_-^\text{GTP} /k_+^\text{GTP} $. Based on these observations, infra-affinity can only be achieved when $k_\pm^\text{GDP} < k_\pm^\text{GTP}$ (see Appendix B and Fig.~\ref{fig:GTPases_bound}). } 
\label{fig:cofactors}
\end{figure}

The cofactors of Hsp70 are needed to reach affinity beyond the equilibrium range. In the case of ultra-affinity, our optimization showed that, first, hydrolysis must be much faster than nucleotide exchange in the substrate-bound state, whereas nucleotide exchange must be much faster than hydrolysis in the absence of a substrate. Second, these reactions must be much faster than substrate binding and unbinding kinetics. These two key requirements are necessary to optimally tune the effective dissociation constant beyond the equilibrium restrictions. Most remarkably, J-proteins and nucleotide exchange factors (NEFs) have a similar role in the Hsp70 system \cite{kamp10}. First, J-proteins bind to a specific sequence of amino acids present in non-native (misfolded) proteins and catalyze the hydrolysis reaction by four order of magnitudes over the basal rate \cite{lauf99}. The second key elements are the NEFs. They have high affinity for the $\mathrm{H_{ADP}}$ state and catalyze the dissociation of ADP \cite{brac15}. NEFs should ideally only boost nucleotide exchange without bound substrate. Nevertheless, J-proteins catalyze hydrolysis much stronger than NEFs catalyze nucleotide exchange in the presence of a substrate, thus, favoring hydrolysis over nucleotide exchange in that case. In the absence of a substrate, hydrolysis is slow (not catalyzed), whereas nucleotide exchange is catalyzed by NEFs (see Fig.~\ref{fig:cofactors}). Experimental observations on the Hsp70 system, thus, match the kinetics requirements found during our optimization.  

Hsp70 can tune its affinity depending on the stress level. Using our model, we can show the effect of stress on the affinity induced by the response of cofactors and the change in nucleotide levels measured during heat-shock experiments. During heat-shock experiments, where the temperature is increased from $37^\circ$ to $45^\circ$, the effect of J-proteins on hydrolysis becomes stronger ($\approx 100\omega_\pm^{S}$) and the effect of NEFs becomes weaker ($\approx 0.1\kappa_\pm, 0.1\kappa_\pm^\text{S}$) \cite{sieg06,grim01}. Moreover, the ATP concentration decreases and the ADP concentration increases leading to a weaker driving force $\Delta\mu$ \cite{soin05,lill84,find83}. We consider a system tuned in the equilibrium range at $\varDelta\mu=10$, which is represented at $\text{H}_\text{normal}$ in Fig.~\ref{fig:bound}. We model the response of cofactors to stress by tuning the hydrolysis in the substrate-bound state by a factor 100 (rates $\omega_\pm^\text{S}$) and decreasing the nucleotide exchange rates by a factor 10 (rates $\kappa_\pm, \kappa_\pm^\text{S}$). In addition, we model the decrease of the ATP ($\text{[ATP]}_\text{stress} \approx 0.4\text{[ATP]}_\text{normal}$) leading to a decrease of the rates $\kappa_+, \kappa_+^\text{S}$ by a factor $e^{-1}$, and the increase of ADP ($\text{[ADP]}_\text{stress} \approx 7\text{[ADP]}_\text{normal}$) leading to an increase of the rates $\kappa_-, \kappa_-^\text{S}$ by a factor $e^{2}$. We obtain a stressed system at $\text{H}_\text{stress}$ in Fig.~\ref{fig:bound}  which can achieve ultra-affinity at $\varDelta\mu=7$ despite a weaker non-equilibrium driving. This example shows that ultra-affinity can dynamically be turned on and off. 

J-domain containing proteins have specific substrate preferences, necessary to recruit Hsp70s on the targets that need their intervention. Since ultra-affinity is triggered by ATP hydrolysis, which in turn is stringently accelerated by the interaction with the J-domain (see thick red arrows $\omega_\pm^\text{M}$ in Fig.~\ref{fig:cofactors}A), ultra-affinity only works on the substrates that have been selected for Hsp70 by J-proteins. In the case of protein folding, energy consumption enhances the affinity of Hsp70s only toward non-native (misfolded, unfolded and aggregated) proteins, because the associated J-proteins have a low affinity for native polypeeptides. After hydrolysis, J-proteins are expelled and unfolding can occur in the ADP state \cite{kamp10,kell14}. After unfolding, the protein is then rejected through nucleotide exchange and unbinds from the ATP state \cite{yuji13,fink16}. The specificity of J-proteins is thus central for efficient protein refolding and allows binding with high affinity and rejecting the substrate after unfolding. Ultra-affinity can only be achieved when reactions induce cycling in the counter clockwise direction as shown in the lower left panel in Fig.~\ref{fig:cofactors}A  (see also discussion from Appendix A). For a partially folded substrate X, nucleotide exchange reactions do not induce directed cycling, which allows the substrate to be rejected with a higher effective dissociation constant $K_\text{eff}$. In an ideal refolding scheme, partially folded substrate should be rejected with infra-affinity, where reactions should induce cycling in the opposite direction of ultra-affinity. In the case of Hsp70, this would require a strongly catalyzed hydrolysis reaction in the absence of substrate. Hence, due to the allosteric barrier $\varDelta\mu>(1/\beta)\ln (K_\mathrm{D}^\text{ATP}/K_\mathrm{D}^\text{ADP})$, as shown in Fig.~\ref{fig:bound}, it may be quite challenging to realize infra-affinity for the Hsp70 system.

Small GTPases, however, may provide a quite similar kinetic scheme for which infra-affinity will be biologically more relevant. Small GTPases are molecular switches which are key regulators in many cellular processes \cite{cher11}. They are GTP-driven machines going trough a GTP hydrolysis cycle similar to Hsp70. Small GTPases also work with two cofactors: GTP hydrolysis is boosted by GTPase activating proteins  (GAPs) and GTP exchange is catalyzed by Guanine nucleotide exchange factors (GEFs) \cite{cher13,voss16}, see Fig.~\ref{fig:cofactors}B. They rely on allosteric regulation, where the active GTP-state binds more tightly to its effector (substrate) than the inactive GDP-state contrary to Hsp70 \cite{herr95,berg09,burg08,junu04}. Surprisingly, bound complexes have a short lifetime \cite{sydo98}.  Infra-affinity with fast binding kinetics can optimize the transmission of signal with fast activation and efficient release to further molecules \cite{herr03,barr13}. In Fig.~\ref{fig:cofactors}B, we show the kinetic requirements to achieve infra-affinity in small GTPases. First, nucleotide exchange must be much faster than hydrolysis in the substrate-bound state and hydrolysis must be much faster than nucleotide exchange in the absence of a substrate. Second, these reactions must be much faster than substrate binding and unbinding kinetics. Most remarkably, GAPs and GEFs fulfills these conditions for small GTPases. GEFs catalyze the dissociation of GDP similar to NEFs, which results in faster nucleotide exchange $\kappa_\pm, \kappa_\pm^\text{E}$ reactions. Contrary to Hsp70, GAPs boost hydrolysis by binding to small GTPases as an effector (substrate), therefore, resulting in faster hydrolysis $\omega_\pm$ reactions \cite{cher13}. Second, these reactions are strongly catalyzed \cite{shut06,voss16}. Experimental observations of small GTPases, thus, match the kinetic requirements for infra-affinity found during our optimization; see Appendix B for more details.

\section*{DISCUSSION AND SUMMARY}

In this work, we have assessed the thermodynamic bounds of ultra-affinity, namely, how effectively the energy available from ATP hydrolysis can be converted into the non-equilibrium enhanced affinity of Hsp70 for its substrates. 

Ultra-affinity can prevent aggregation and improve the refolding efficiency. Most substrates bound to Hsp70 are protected and do not aggregate \cite{sekh15}. In addition, substrates bound to Hsp70 are unfolded rather than misfolded compared to free specimens \cite{shar10,cler15}. The unfolding process could be due to an interaction between Hsp70 and its substrate in the closed ADP-state \cite{maye13}. Therefore, this unfolding activity coupled with ultra-affinity could ultimately help to shift the energy landscape to favor the refolding of the substrate after release in an unfolded, refoldable species. Recent single-molecule experiments show new insights into the role of Hsp70 in protein folding \cite{mash16}. For instance, they show that Hsp70 also protects partially folded structures against aggregation in addition to misfolded substrates. Moreover, they find that Hsp70 can both stabilize and destabilize native structures depending on the nucleotide concentrations. It would be interesting to add protein dynamics to our model (as sketched in Fig.~\ref{fig:cofactors}A) and investigate refolding strategies under different stress levels. 

The relation between stress and affinity could be further investigated in experiments. In Fig.~\ref{fig:4state}, we show that changes in nucleotide levels \cite{soin05,lill84,find83} and temperature dependent cofactors \cite{sieg06,grim01} can lead to an increase of affinity with a smaller driving force $\Delta\mu$. Using our framework, we explain how these two responses to stress lead to a higher affinity for substrates (see Fig.~\ref{fig:cofactors}). It would be interesting to make in vitro experiments investigating the relation between the driving force $\Delta\mu$, the individual ADP and ATP levels together with the effective affinity. Adjusting the nucleotide levels accordingly, increasing the concentration of J-proteins and decreasing the concentration of NEFs should trigger an ultra-affinity response in Hsp70, thus, mimicking a heat-shock without any temperature change.

We found that infra-affinity is quite difficult to realize with Hsp70 due to an allosteric barrier. For small GTPases, however, we show how infra-affinity can be essential for efficient molecular switches. While the link between infra-affinity and small GTPases has not been experimentally observed yet, the function of small GTPases can benefit from infra-affinity. We have shown under which conditions small GTPases could achieve fast binding and unbinding with an effector coupled with low affinity (see Fig. \ref{fig:cofactors}B and Appendix B). Our analysis was based on biological properties of cofactors and the fact that the GTP state has a higher affinity to effectors than the GDP state. Our framework shows that infra-affinity could only be achieved if the binding and unbinding kinetics in the GTP state are faster than the GDP state ($k^\text{GTP}_\pm > k^\text{GDP}_\pm$) which has not been confirmed yet experimentally to our knowledge. Measuring slow binding and unbinding rates in a low affinity state may be difficult during experiments. Nevertheless, if an experiment directly measured an effective dissociation constant $K_\text{eff}$ lower than the equilibrium range, our framework could be used to give insights into the the binding and unbinding kinetics in the GDP state.

The framework proposed in \cite{rios14} and exploited here might also be applicable to other non-equilibrium systems relying on allosteric regulation. Specifically, other chaperones system such as Hsp90s, Hsp100s or the GroEL-GroES system \cite{yuji13,fink16}, which have been proposed to exhibit non trivial non-equilibrium properties \cite{bard15} could benefit from our general scheme and give us more insights into the role of energy consumption in these systems.

\appendix
\section*{Appendix A: Derivation of the thermodynamic bounds} \label{sec:Appendix}
\begin{figure}
\centering
\includegraphics[width=.3\linewidth]{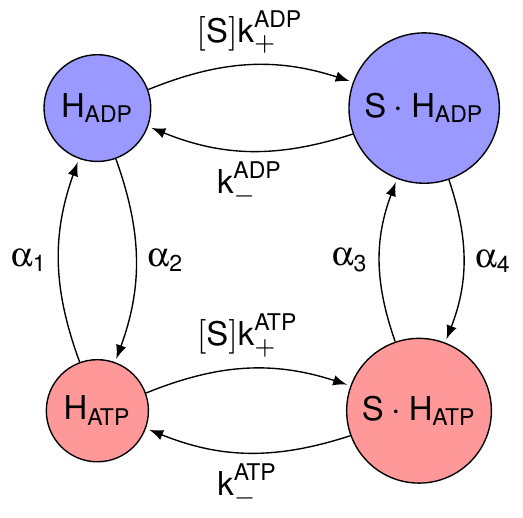}
\caption[Four state model]{\label{fig:4state_simpl} Four state model from Fig.~\ref{fig:4state} with total transition rates $\alpha_1,\ldots,\alpha_4$ including both hydrolysis and nucleotide exchange reactions.
}
\end{figure}

We introduce a simple binding model in Fig.~\ref{fig:4state_simpl}, where we consider the total rates $\alpha_1=\omega_++\kappa_-$, $\alpha_2=\omega_-+\kappa_+$, $\alpha_3=\omega_+^S+\kappa_-^S$, and $\alpha_4=\omega_-^S+\kappa_+^S$, which include both hydrolysis $\omega_{\pm}$ and nucleotide exchange $\kappa_{\pm}$ reactions. We define the coarse-grained affinity
\begin{equation}\label{gen_aff}
  \widetilde{\mathcal{A}}\equiv
  \ln \frac{\alpha_2 \alpha_3 k^\textrm{ADP}_- k^\textrm{ATP}_+}{\alpha_1 \alpha_4 k^\textrm{ADP}_+ k^\textrm{ATP}_-}= \ln \frac{K_{\text{D}}^\textrm{ADP} \left(\omega_- + \kappa_+ \right) \left(\omega_+^S + \kappa_-^S \right)}{K_{\text{D}}^\textrm{ATP} \left(\omega_+ + \kappa_- \right) \left(\omega_-^S +\kappa_+^S \right)},
\end{equation}
where we have identified the total transition rates in the second step. Note that this coarse-grained affinity $\widetilde{\mathcal{A}}$ should not be confused with the chemical affinity which is the inverse of the effective dissociation constant $K_\textrm{eff}$. A positive sign of $\widetilde{\mathcal{A}}$ indicates a cycling in the counter clockwise direction. Most importantly, $\widetilde{\mathcal{A}}\neq0$ can be attained only if the system is driven out of equilibrium. The local detailed balance relation from Eqs.~\ref{eq:LDB_substrate} and \ref{eq:LDB}
impose the following constraints on the transition rates
\begin{align} \label{cond1}
\frac{\omega_+ \kappa_+}{\omega_- \kappa_-} = \frac{\omega_+^S \kappa_+^S}{\omega_-^S \kappa_-^S}   &= e^{\beta\varDelta\mu},\\
\label{cond2}
\frac{\omega_+ \kappa_+^S K_{\text{D}}^\textrm{ATP}}{\omega_- \kappa_-^S K_{\text{D}}^\textrm{ADP}} &= e^{\beta\varDelta\mu},\\
\label{cond3}
\frac{\omega_+^S \kappa_+ K_{\text{D}}^\textrm{ADP}}{\omega_-^S \kappa_- K_{\text{D}}^\textrm{ATP}}& = e^{\beta\varDelta\mu},
\end{align}
where $\beta=1/(\kb T)$ is the inverse thermal energy.
From Eqs.~\ref{cond1}-\ref{cond3}, the coarse-grained affinity \ref{gen_aff} is bound between
\begin{equation}\label{driv_bound}
  -\beta\varDelta\mu \leq  \widetilde{\mathcal{A}}  \leq \beta\varDelta\mu.
\end{equation}
Thereby, a maximization over $\kappa_\pm,\kappa_\pm^S,\omega_\pm,\omega_\pm^S$ while keeping
Eqs.~\ref{cond1}-\ref{cond3} fixed is equivalent to a maximization over the coarse-grained rates $\alpha_1,\ldots,\alpha_4$ while satisfying Eq.~\ref{driv_bound}.

Denoting the steady state probabilities by $P_i$, where $i$ labels the states $i=\mathrm{H_{ATP}},\mathrm{H_{ADP}},\mathrm{S\cdot H_{ ATP}},\mathrm{S\cdot H_{ADP}}$, allows us to write the effective dissociation constant in the form
\begin{equation} \label{Keff}
 K_\textrm{eff} = \frac{P_{\text{off}}}{P_{\text{on}}}[\text{S}] \equiv \frac{P_{\text{H}_{\text{ADP}}} + P_{\text{H}_{\text{ATP}}}}{P_{\mathrm{S\cdot \text{H}_{ADP}}} +
P_{S\cdot\text{H}_\mathrm{ATP}}}[\text{S}],
\end{equation}
which is the inverse of the effective affinity for substrates. We calculate the stationary probability distribution with standard methods \cite{schn76,hill05} and obtain
\begin{equation}
K_\textrm{eff}=
\frac{(\alpha_1+\alpha_2)(\alpha_3k^\textrm{ADP}_-+\alpha_4k^\textrm{ATP}_-)+B}{(\alpha_3+\alpha_4)(\alpha_1k^\textrm{ADP}_++\alpha_2k^\textrm{ATP}_+)+C}
,\label{Kneq}
\end{equation}
where $B=(\alpha_1+\alpha_2)k^\textrm{ADP}_-k^\textrm{ATP}_-+(\alpha_3k^\textrm{ADP}_-k^\textrm{ATP}_++\alpha_4k^\textrm{ADP}_+k^\textrm{ATP}_-)[S]$ and $C=\alpha_1k^\textrm{ATP}_-k^\textrm{ADP}_++\alpha_2k^\textrm{ADP}_-k^\textrm{ATP}_++(\alpha_3+\alpha_4)k^\textrm{ADP}_+k^\textrm{ATP}_+ [S]$. The terms $B$ and $C$ are linear functions of the total transition rates $\alpha_1,\ldots,\alpha_4$ which do not allow the dissociation constant to be controlled beyond the equilibrium restrictions, since $ \min [K_\mathrm{D}^\text{ADP},K_\mathrm{D}^\text{ATP}]\le B/C\le  \max [K_\mathrm{D}^\text{ADP},K_\mathrm{D}^\text{ATP}]$. Note that in the limit of high substrate concentration, the effective dissociation constant Eq. \ref{Kneq} is also restricted to the equilibrium range.

Any dissociation constant can be written in the form
\begin{equation}
 K_\textrm{eff}=
 \frac{p k^\textrm{ADP}_-+(1-p)k^\textrm{ATP}_-}{q k^\textrm{ADP}_++(1-q)k^\textrm{ATP}_+}
\label{eq:Keff_pq}
 \end{equation}
where $p,q$ are positive weight parameters satisfying $0\le p,q\le1$ and
\begin{equation} 
 \e^{-\beta\varDelta\mu}\le\frac{p(1-q)}{(1-p)q}\frac{k_-^\mathrm{ADP}k_+^\mathrm{ATP}}{k_+^\mathrm{ADP}k_-^\mathrm{ATP}}\le\e^{\beta\varDelta\mu},
 \label{eq:pqConstraints}
\end{equation}
which follows from Eqs.~\ref{gen_aff} and \ref{driv_bound}. The maximal attainable range for the dissociation constant
 is given by
 \begin{equation} \label{eq:Kbounds}
  K_\text{D}^\text{min}\le K_\text{eff}\le K_\text{D}^\text{max}
 \end{equation}
where $ K_\text{D}^\text{min}$ ($ K_\text{D}^\text{max}$) is the minimum (maximum) of Eq.~\ref{eq:Keff_pq} with respect to $p,q$ within the allowed range from Eq.~\ref{eq:pqConstraints}. The effective dissociation constant $K_\textrm{eff}$ is best controlled if the vertical transitions $\alpha_1,\ldots,\alpha_4$ are much faster than the transition rates involving substrate binding and unbinding, where $p\approx \alpha_3/(\alpha_3+\alpha_4)$ and $q\approx\alpha_1/(\alpha_1+\alpha_2)$.
Note that $K_\textrm{eff}$ saturates to its lower (upper) limit in Eq.~\ref{eq:Kbounds} if the weight parameters $p,q$ are chosen such that the expression in Eq.~\ref{eq:pqConstraints} equals $\e^{\beta\varDelta\mu}$ ($\e^{-\beta\varDelta\mu}$), i.e., Eq.~\ref{eq:pqConstraints} must saturate as well.

Specifically, for Hsp70 with the binding and unbinding kinetics from Table~\ref{4state_realrate}, we obtain the following bounds. For $\varDelta\mu\ge0$ the minimal dissociation constant is given by
\begin{equation}
 K_\text{D}^\text{min} =
\dfrac{K^\textrm{ADP}_D K^\textrm{ATP}_D\left(k^\textrm{ATP}_+-k^\textrm{ADP}_+ e^{\beta\varDelta\mu }\right)^2\sqrt{\Theta_+}}
{\left(\left(k^\textrm{ATP}_- e^{\beta\varDelta\mu }-k^\textrm{ADP}_- e^{\beta\varDelta\mu } \right) \left(k^\textrm{ADP}_+-k^\textrm{ATP}_+\right)+\sqrt{\Theta_+}\right)
\left(\left(k^\textrm{ADP}_+ k^\textrm{ATP}_- e^{\beta\varDelta\mu }-k^\textrm{ADP}_- k^\textrm{ATP}_+\right) \left(e^{\beta\varDelta\mu }-1\right)-\sqrt{\Theta_+}\right)},
\end{equation}
where
\begin{equation}
\Theta_+ =\left(k^\textrm{ADP}_- - k^\textrm{ATP}_-\right) \left(k^\textrm{ADP}_+ - k^\textrm{ATP}_+\right) \left(-1 + e^{\beta\varDelta\mu}\right)  \left(e^{\beta\varDelta\mu} - \frac{K^\textrm{ADP}_D}{K^\textrm{ATP}_D} \right) e^{\beta\varDelta\mu}k^\textrm{ADP}_+ k^\textrm{ATP}_-.
\end{equation}
Note that the system is optimally tuned for ultra-affinity when $\widetilde{\mathcal{A}}=\beta\varDelta\mu$ (i.e., $p,q$ maximize Eq.~\ref{eq:pqConstraints}). Therefore, the lowest dissociation constant requires hydrolysis to be much faster than nucleotide exchange in the substrate-bound state, whereas in the absence of a substrate nucleotide exchange must be much faster than hydrolysis.

Notably, infra-affinity requires a large enough chemical potential $\varDelta\mu>(1/\beta)\ln(K^\mathrm{ATP}_\mathrm{D}/K^\mathrm{ADP}_\mathrm{D})$, in which case the maximal dissociation constant
is given by
\begin{equation}
 K_\text{D}^\text{max} =
\dfrac{K^\textrm{ADP}_D K^\textrm{ATP}_D\left(k^\textrm{ADP}_+-k^\textrm{ATP}_+ e^{\beta\varDelta\mu }\right)^2\sqrt{\Theta _-}}
{\left(\left(k^\textrm{ADP}_- e^{\beta\varDelta\mu } -k^\textrm{ATP}_- e^{\beta\varDelta\mu } \right) \left(k^\textrm{ADP}_+-k^\textrm{ATP}_+\right)+\sqrt{\Theta _-}\right)
   \left(\left(k^\textrm{ADP}_- k^\textrm{ATP}_+ e^{\beta\varDelta\mu }-k^\textrm{ADP}_+ k^\textrm{ATP}_-\right) \left(e^{\beta\varDelta\mu }-1\right)+\sqrt{\Theta _-}\right)},
\end{equation}
where
\begin{equation}
\Theta_- = \left(k^\textrm{ADP}_--k^\textrm{ATP}_-\right) \left(k^\textrm{ADP}_+-k^\textrm{ATP}_+\right) \left(e^{\beta\varDelta\mu }-1\right) \left(e^{\beta\varDelta\mu }-\frac{K^\textrm{ATP}_D }{K^\textrm{ADP}_D}\right)e^{\beta\varDelta\mu } k^\textrm{ADP}_- k^\textrm{ATP}_+.
\end{equation}
However, small chemical potentials $0\le\beta\varDelta\mu\le\ln(K^\mathrm{ATP}_\mathrm{D}/K^\mathrm{ADP}_\mathrm{D})$ do not allow infra-affinity, since $K_\text{D}^\text{max}=K^\mathrm{ATP}_\mathrm{D}=k_-^\mathrm{ATP}/k_+^\mathrm{ATP}$.
Note that the system is optimally tuned for infra-affinity when $\widetilde{\mathcal{A}}=-\beta\varDelta\mu$ (i.e., $p,q$ minimize Eq.~\ref{eq:pqConstraints}). Therefore, the maximal dissociation constant requires nucleotide exchange to be much faster than hydrolysis in the substrate-bound state, whereas in the absence of a substrate hydrolysis must be much faster than nucleotide exchange.

\section*{Appendix B: Analysis of Small GTPases} \label{sec:AppendixB}
\begin{figure}
\centering
\includegraphics[width=.7\linewidth]{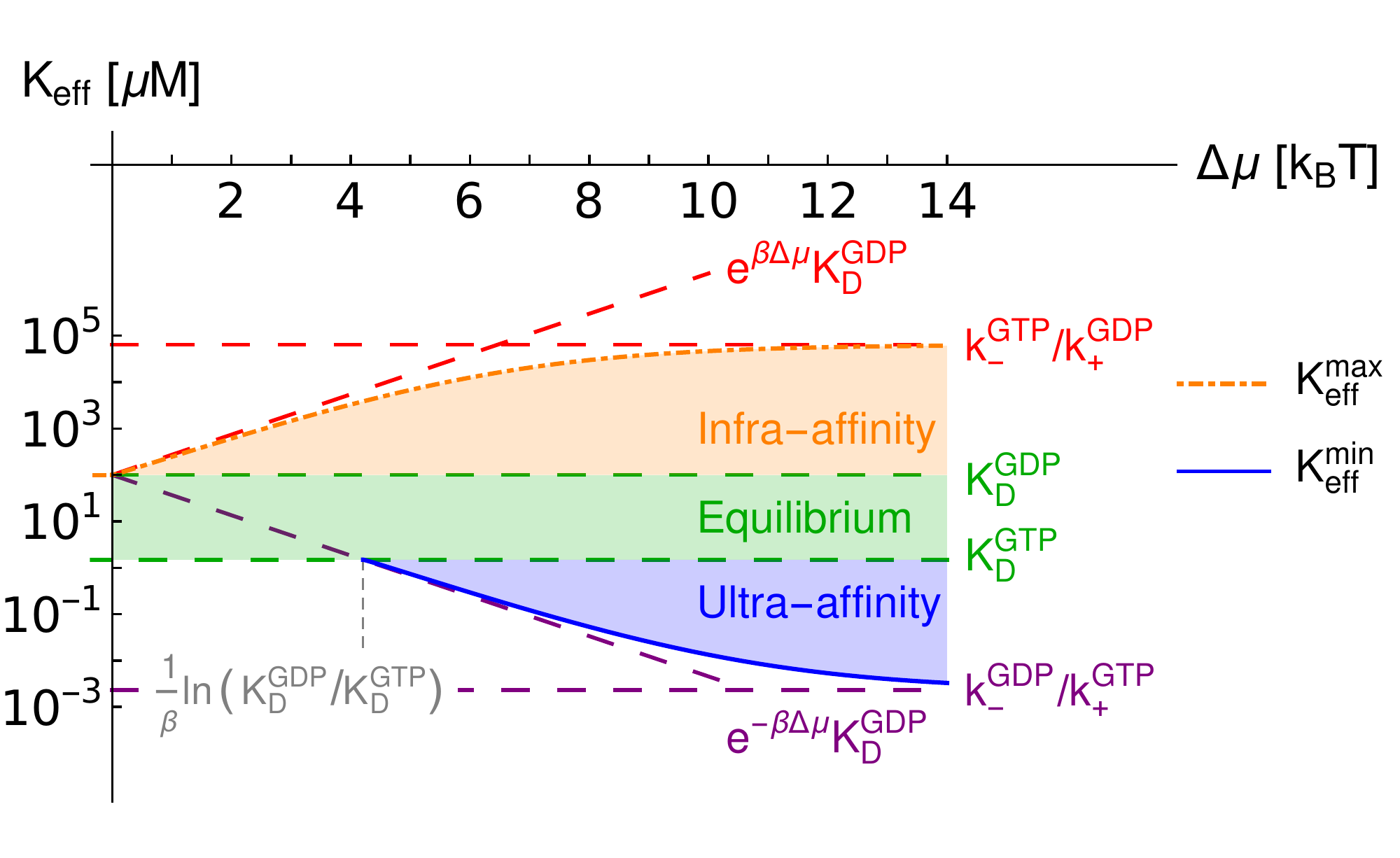}
\caption{Thermodynamic bounds on the effective dissociation constant for small GTPases based on experimental binding rates. In the GTP state $k^\textrm{GTP}_+ = 4.3 \text{s}^{-1}\mu\text{M}^{-1} $, $k^\textrm{GTP}_- = 6.4\text{s}^{-1}$ where $K_\mathrm{D}^\text{GDP} = 1.5\mu\text{M}$ \cite{berg09}. To achieve infra-affinity, the binding and unbinding kinetics must be slow, we choose rate $k^\textrm{GDP}_+ = 10^{-4} \text{s}^{-1}\mu\text{M}^{-1} $ and $k^\text{GDP}_- = 10^{-2} \text{s}^{-1}$ such that $K_\mathrm{D}^\text{GDP} \approx 10^2\textup{--}10^3 \, K_\mathrm{D}^\text{GTP}$ and matches experimental estimation \cite{berg09,burg08,junu04}. Ultra-affinity can be achieved only with a minimum driving force $\varDelta\mu > (1/\beta)\ln (K_\mathrm{D}^\text{GDP}/K_\mathrm{D}^\text{GTP})$.}
\label{fig:GTPases_bound}
\end{figure}

We provide in this section a detailed analysis of small GTPases. Small GTPases go through a GTP hydrolysis cycle similar to Hsp70. Our four state model Fig.~\ref{fig:4state_simpl} introduced previously can be extended to small GTPases by changing ATP(ADP) to GTP(GDP). Small GTPases have fast binding kinetics and high affinity to substrate in their active GTP state. In the inactive GTP state, they have low affinity for substrate and slow binding kinetics \cite{berg09,burg08,junu04}. Experiments have measured the binding and unbinding rates in the GTP state \cite{berg09}, however, only estimation of binding kinetics in the GDP state are available at the current time. We show here under which kinetic conditions small GTPases can be tuned for infra-affinity. Small GTPases work with cofactors similar to Hsp70. Therefore, we can tune the effective dissociation constant $K_\textrm{eff}$ by varying the the hydrolysis $\omega_\pm$ and nucleotide exchange rate for a given energy budget $\Delta \mu$. Similar to Appendix A, we extremize Eq.~\ref{eq:SmallGTPpqConstraints} with respect to $p,q$ within the allowed range from
\begin{equation} 
 \e^{-\beta\varDelta\mu}\le\frac{p(1-q)}{(1-p)q}\frac{k_-^\mathrm{GDP}k_+^\mathrm{GTP}}{k_+^\mathrm{GDP}k_-^\mathrm{GTP}}\le\e^{\beta\varDelta\mu},
 \label{eq:SmallGTPpqConstraints}
\end{equation}
which is adopted from Eq.~\ref{eq:pqConstraints}, where the kinetic parameters from Fig.~\ref{fig:4state} are replaced by the ones from Fig.~\ref{fig:cofactors}B.   We obtain bounds for $K_\textrm{eff}$ as shown in Fig.~\ref{fig:GTPases_bound}. Similar to Hsp70, simple lower and upper bounds are given respectively by $\min \left[e^{\beta\varDelta\mu}K_\mathrm{D}^\text{GDP}, k^\textrm{GTP}_-/k^\textrm{GDP}_+ \right]$ and $\max \left[e^{\beta\varDelta\mu}K_\mathrm{D}^\text{GDP}, k^\textrm{GTP}_-/k^\textrm{GDP}_+ \right]$. 

\section*{AUTHOR CONTRIBUTIONS}
B.N. performed research; B.N., D.H., U.S. and P.D.L.R. designed research and wrote the article.
\section*{ACKNOWLEDGMENTS}
We thank Alessandro Barducci for helpful scientific discussion. P.D.L.R thanks the Swiss National Science Foundation for financial support under the grant number 200020\_163042.


%
%
%



\end{document}